%% file: main.tex
\documentclass[conference]{IEEEtran}
\IEEEoverridecommandlockouts
\usepackage{cite}
\usepackage{amsmath,amssymb,amsfonts}
\usepackage{algorithmic}
\usepackage{graphicx}
\usepackage{textcomp}
\usepackage{xcolor}

\def\BibTeX{{\rm B\kern-.05em{\sc i\kern-.025em b}\kern-.08em
    T\kern-.1667em\lower.7ex\hbox{E}\kern-.125emX}}
\begin{document}


\input{./subsection/000_title.tex}

\input{./subsection/010_abstract.tex}

\section{Introduction}\label{S_introduction}
\input{./subsection/020_introduction.tex}

\section{BUILD TIME OF MODERN SOFTWARE}\label{S_buildtime}

\input{./subsection/030_build_time.tex}

\section{Design and Implementation}\label{S_design}
\input{./subsection/050_design_implement.tex}

\section{Evaluation}\label{S_evaluation}
\input{./subsection/060_evaluation.tex}
 
\section{Related Work}\label{S_relatedwork}
\input{./subsection/070_related_work.tex}

\section{Conclusion}\label{S_conclusion}
\input{./subsection/080_conclusion.tex} 
 
\section*{Acknowledgment}
\input{./subsection/090_acknowledgment.tex}

\input{./subsection/095_reference.tex} 
 
\end{document}

%% file: subsection/000_title.tex
\title{TAOS-CI: Lightweight \& Modular Continuous Integration System for Edge Computing }

\input{./subsection/000_020_author_type1.tex}

\maketitle


%% file: subsection/000_020_author_type1.tex
\author{

\IEEEauthorblockN{
Geunsik Lim,
MyungJoo Ham,
Jijoong Moon,
Wook Song,
Sangjung Woo, and
Sewon Oh 
} 

\IEEEauthorblockA{
Artificial Intelligence Center, Samsung Research, Seoul, South Korea 
}

\{geunsik.lim, myungjoo.ham,  jijoong.moon, wook16.song, sangjung.woo, sewon.oh\}@samsung.com\


\thanks{Corresponding author: MyungJoo Ham, myungjoo.ham@samsung.com}
}

%% file: subsection/010_abstract.tex
\begin{abstract}
With the proliferation of IoT and edge devices, we are observing a lot of consumer electronics becoming yet another IoT and edge devices.
Unlike traditional smart devices, such as smart phones, consumer electronics, in general, have significant diversities with fewer number of devices per product model.
With such high diversities, the proliferation of edge devices requires frequent and seamless updates of consumer electronics, which makes the manufacturers prone to regressions because the manufacturers have less resource per an instance of software release; i.e., they need to repeat releases by the number of product models times the number of updates.
Continuous Integration (CI) systems can help prevent regression bugs from actively developing software packages including the frequently updated device software platforms.
The proposed CI system provides a portable and modular software platform automatically inspecting potential issues of incoming changes with the enabled modules: code format and style, performance regressions, static checks on the source code, build and packaging tests, and dynamic checks with the built binary before deploying a platform image on the IoT and edge devices.
Besides, our proposed approach is lightweight enough to be hosted in normal desktop computers even for dozens of developers.
As a result, it can be easily applied to a lot of various source code repositories.
Evaluation results demonstrate that the proposed method drastically improves plug-ins execution time and memory consumption, compared with methods in previous studies.
\\
\end{abstract}

\begin{IEEEkeywords}
continuous integration, continuous test, software regression, platform build, code review
\end{IEEEkeywords}

\let\thefootnote\relax\footnote{This research was supported by Rocky 2018 (RAV0117ZZ-82RF), the AI Center project of Samsung Research. This work is supported in part by the AI Core Platform (RAJ0118ZZ-CBRF) through AI Center of Samsung Research.}

%% file: subsection/020_introduction.tex
 Recently, the computing environment has evolved from cloud computing to edge computing. In edge computing, Internet-of-Things (IoT) devices store the collected data on the cloud computer in real time. Many existing devices are equipped with connectivity such as the internet. As a result, various IoT devices with different hardware structures have been increasing exponentially.

Such IoT devices usually operate continuously 24/7. To guarantee uninterrupted service, such devices have to be updated frequently and automatically\textemdash we cannot afford manually updating IoT devices or delaying updates for a long time\textemdash to help prevent potential security defect, performance deterioration, or behavioral issues in run time.

The complexity of software is becoming higher even for lightweight IoT devices. With such a complexity, a software update incurred by a code commit is required to be integrated to deployed system \cite{6448697} in real time in order to reduce the latencies, which is critical to the security and overall system robustness: we may want every node to share the same recent versions. Therefore, the software integration \cite{6802994} and deployment latency from the code commit is emerging as a key pain point for software platform deployment.

Especially, users continuously require new features of the software and there are large software packages with more than hundreds of actively participating developers in a single software package such as \textit{Caffe}, and \textit{Tensorflow} in case of Edge Artificial Intelligence (AI). The more software size gets bigger and complicated, the more we need to inspect the code and prevent errors from new code commits before deployment especially if we have a lot of devices deployed with diversities in their architectures and hardware configurations; more devices mean more potential failures. \cite{6606557} Therefore, both the integration and deployment latencies \cite{Waller:2015:IPB:2735399.2735416} and the error prevention capabilities of CI systems \cite{6802994} are critical.

The reminder of the paper is organized as follows. The build time of modern software is described in Section II. Section III addresses the design and implementation of the proposed techniques in detail. Section IV shows the evaluation results, and related work is described in Section V. Finally, Section VI concludes the paper.

%% file: subsection/030_build_time.tex
\begin{figure}
\centering
\includegraphics[width=0.99\columnwidth,height=2.0in]{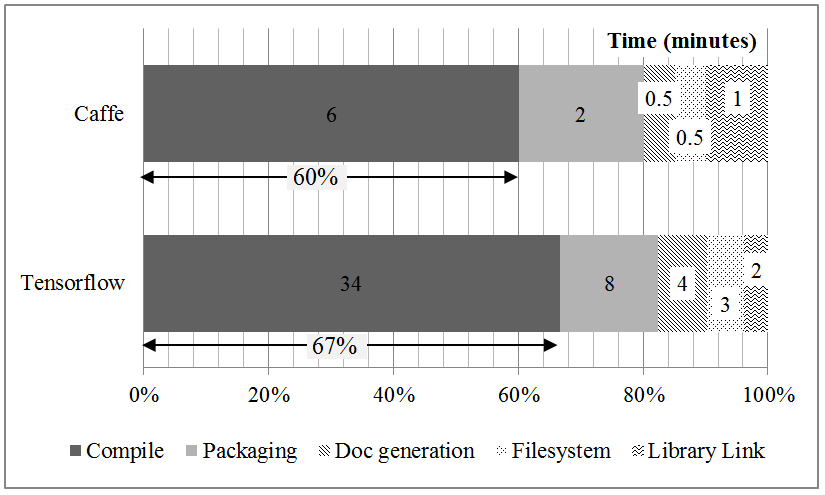}
\caption{Breakdown of build operation}
\label{fig:breakdown_of_build_operation}
\end{figure}

Software becomes sophisticated due to its growing functionality and code complexity. As a result, build and integration latency \cite{8115619} are increasing as well. Figure~\ref{fig:breakdown_of_build_operation} shows the time cost of building \cite{Mcconnell:1996:DBS:624614.625626} a well-known deep-learning frameworks, \textit{Caffe} and \textit{Tensorflow}. Compile latency of \textit{Tensorflow} accounts for 67\% (34 minutes) of the total 51 minutes. Compile latency of \textit{Caffe} accounts for 60\% (6 minutes) of the total 10 minutes. Over the half of build time is consumed by compilation and incorrect source codes might cause the longer build time. Therefore providing correct code is more important to save resources.

To filter out many code commits that may incur errors and failures before actually building them in the build server, another operation must be designed to check the correctness of the given code commit \cite{Yu2016} without actually building it. We should also check the format of the given code commit \cite{7180096} if it complies with the given rules. It is critical to secure a software code management system technology that checks whether the built code is valid in short time.

Note that building a software package in a build system\textemdash not in a workstation of an individual developer\textemdash of a CI system cannot utilize the benefit of incremental build in general. The depending libraries and toolchains keep being updated continuously; i.e., several times a day in large development project.

Besides, for the integrity of resulting binaries guaranteeing that the binaries are bit-to-bit the same if the code set is identical regardless of which server is used to build or which OS is used to host the build tasks, providing clean VMs or environments for each and every build task is recommended: e.g., GBS (\textit{Tizen}), Open Build Service (\textit{openSUSE}), and pbuilder/pdebuild (\textit{Ubuntu/Debian}).

In addition, we need to inspect the overall integrity before the resulting binary packages from the new source code is actually integrated into a system so that the updated platform binary image can be deployed quickly to various IoT devices without disrupting the devices. The following section defines elements that impede the ongoing integration \cite{10.1007/978-0-387-09684-1_23} of software code. Besides, we depict our proposed techniques to minimize a hurdle of continuous integration.

%% file: subsection/050_design_implement.tex
In this section, we describes how to minimize the non-productive situations that reviewers should repeatedly check to ensure that source code is well written. In addition, the proposed system performs (1) the format checker (before the build) and (2) the audit checker (after the build) by separating the source code. We suggest a novel system that be reassembled with lots of user-defined check modules to improve the integration speed of source code. 

Software updates have to be applied to the edge computing environment in real time. Software packages that are created by tens or hundreds of developers must continuously integrate the source code. Especially, a code reviewer finds repeatedly software defects from the submitted source code. \cite{10.1007/978-0-387-09684-1_23}, \cite{7962384} Our proposed system, \textbf{TAOS-CI}, automatically checks internal defects in the software and reports a result to the developers.

\begin{figure}
\centering
\includegraphics[width=0.99\columnwidth,height=2.4in]{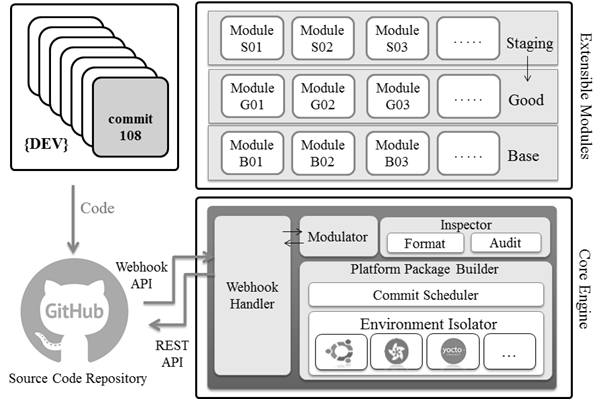}
\caption{System architecture of TAOS-CI}
\label{fig:system_architecture}
\end{figure}

Figure~\ref{fig:system_architecture} shows the overall  flow of TAOS-CI. It consists of two blocks to control a source code repository such as GitHub: (1) the Core Engine to conduct CI processes and (2) Extensible Modules to maintain base, good, and staging group. \cite{6976106}, \cite{6224294}, \cite{5291914}

\subsection{Webhook Handler}
Whenever a developer submits a source code to the collaborative management repository in the cloud environment, the repository sends a webhook event message to the specified development server. The webhook handler interprets the received event message and performs an operation to handle the event. Each time a developer submits a source code to a development server, the webhook handler receives a standardized webhook event message from code management server. Then, the webhook handler delivers the event message to the Inspector via the Modulator.

\subsection{Platform Package Builder}
The source code must be merged into a buildable state. The proposed system selectively supports package build of software platforms such as \textit{Ubuntu} \cite{thomas2006beginning}, \textit{Yocto} \cite{salvador2014embedded}, and \textit{Tizen} \cite{jaygarl2014professional}. It is activated when developers write a packaging script for their software platform. It provides the scalability of platform build by allowing developers to add their build modules if they need to append another software platform.

\begin{itemize}
\item \textbf{Environment Isolator}: Using virtualization technology to build two or more source codes in an isolated environment at the same time, consumes big hardware resources. Therefore, our proposed system recognizes a folder as a root folder designated to be able to build multiple source codes separately. Then \textit{Commit Scheduler} prepares the necessary dependency packages, and then combines the source code into the package to establish the structure for building and testing.

\item \textbf{Commit Scheduler}: \cite{Yu2016} The platform package builder uses two queues such as a wait queue and a run queue to control merge requests of source code. First, the wait queue is designed with the First-in First-Out (FIFO) structure that executes first the job that a first merge request entered the queue. When a job needs to be replaced in the front of the wait queue, it is selected for removal. Next, the run queue contains priority values for each merge request, which will be used by the \textit{commit scheduler}. At this time, it is possible to perform the build in parallel until the number of run queues. Administrator has to configure an appropriate number of a maximum run queue. When the job is successfully finished, the first request jobs waiting in the wait queue are moved to the run queue in order. When a user places duplicated request jobs in the run queue, a \textit{victim maker} automatically kills the oldest jobs from the run queue.
\end{itemize}

\subsection{Modulator}
It supports a plugin structure that allows developers to add and delete extensible facilities if the generated source code is a valid code. At this time, the supported plug-in facilities are classified into the following three types and operate sequentially.
\begin{itemize}
\item \textbf{Plugins-base}: Base modules that are maintained via plug-in and plug-out, but which are mandatorily required to be performed, belong to this group.
\item \textbf{Plugins-good}: Validated source code, well-defined functions, and modules that have passed high stability tests are located into this group.
\item \textbf{Plugins-staging}: Modules that do not have enough stability and functionality to complete but have good feature are in this group. It means modules that can be moved into \textit{Plugins-good} group when the review process and the aging test are completed.
\end{itemize}

\subsection{Inspector}
Basically, the source code files should be independent of the CPU architecture. The source code before compilation should be able to be checked by the Inspector. When the developer submits their source code to the configuration management server, the inspector checks the source code level. Also, it checks for binary files that have been compiled. By dividing the work into two steps before and after compilation, unnecessary resource waste of the continuous integration server \cite{4599493} can be prevented.

\begin{itemize}
\item \textbf{Format}: Run before the Platform Package Builder begins to compile the source. If the source code does not pass the Format step, stop without running the \textit{Platform Package Builder}. And the developer gets the report on errors in the source code.
\item \textbf{Audit}: When all of the Format modules are successfully completed for source code, the source codes enter the Audit phase. The Audit inspector validates the binary code generated after the source code is built.
\end{itemize}

%% file: subsection/060_evaluation.tex
We have experimented with a normal desktop computer to show how our proposed system can run in common personal computers. The testbed is established with the hardware resources that consists of Intel Core i7-5820K Processor (6 Cores with 12 Threads, 3.3GHz, 15MB Cache), Samsung DDR3 16GB, SSD 850 PRO 512GB (MZ-7KE512), Intel 1 Gbps Ethernet Controller. The operating system is installed with Ubuntu 16.04 LTS, Linux 4.4, and 1:1 Linux thread model (NPTL). The CPU policy is set as performance mode to reduce the effects of power management mechanisms. The Intel Ethernet Controller supports a theoretical maximum bandwidth of 1 Gbps. Then, we have installed \textit{Jenkins} (the existing popular system) and TAOS-CI (the proposed system) to verify the effectiveness of the proposed system.

\begin{figure}
\centering
\includegraphics[width=.99\columnwidth,height=2.4in]{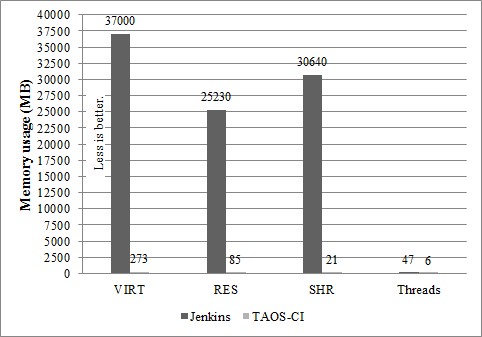}
\caption{Memory usage: VIRT, RES, SHR, and Threads}
\label{fig:memory_usage}
\end{figure}

\subsection{Memory Consumption}

Figure~\ref{fig:memory_usage} shows that the memory consumption is significantly reduced with the proposed system compared to the existing system, \textit{Jenkins}.  The x-axis describes four items as following:
\begin{itemize}
\item \textbf{VIRT}: It means a virtual set size as the  total  amount  of  virtual  memory  used  by the task. It includes all code, data and shared libraries  plus  pages that have  been  swapped out and pages that have been mapped but not used.
\item \textbf{RES}: It means a resident set size. It is the non-swapped physical memory a task has used.
\item \textbf{SHR}: It means a shared memory size as the amount of shared memory used by a task. It simply reflects memory that could be potentially shared with other processes.
\item \textbf{Threads}: It is the number of Linux threads.
\end{itemize}

\textit{Jenkins} has JAVA-based system architecture because it focuses on server-based scalable software architecture. The Java virtual machine based execution structure drastically increases memory usage. 
From our analysis, the result shows that the existing system, \textit{Jenkins}, is executing too many functions, even though those are not used actually. However, note that the proposed system has almost the same degree of portability compared to that of \textit{Jenkins}. The proposed system consists of \texttt{BASH} scripts and \texttt{PHP} scripts, which are available in equally many operating systems.

\subsection{Execution speed of Plug-in Modules}

\begin{figure}
\centering
\includegraphics[width=0.99\columnwidth,height=2.4in]{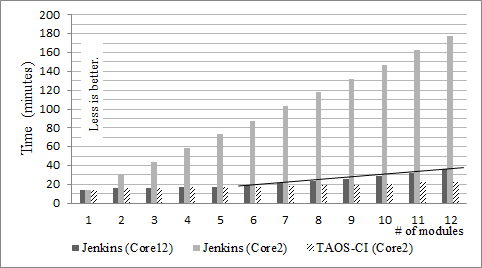}
\caption{Execution time of plug-in modules}
\label{fig:execution_time_of_plugin_modules}
\end{figure}

Figure~\ref{fig:execution_time_of_plugin_modules}  compares the execution latencies when we try to combine different inspection modules. We compared the compile latencies of \textit{Caffe}, one of the popular deep learning frameworks. There are a lot of duplicated code routines in the existing system, \textit{Jenkins}, because there is no unit to manage the modules integrated. In addition, because the most of the modules are not unified, there are lots of duplicated operations among the modules; it is very difficult to work with common and shared operations due to the existing naive design structure. Moreover, the execution speed of the existing system at compile time was directly dependent on the performance of the multicore hardware.

In the existing system, the execution speed of the entire modules can be improved when the number of CPUs can be allocated as many as the number of modules. At this time it has to be executed in parallel. However, the execution speed gradually grows late because the number of increased modules results in I/O cost.

From our analysis, the I/O bottleneck occurs from the point where the number of modules is 6. As a result of that, we could figure out that a naive system design tackles the high performance of Solid-State Drive (SSD) storage.
The proposed system provides a modulator that consistently controls the functions that modules share and use commonly. The modulator helps developers developing new modules by simplifying the implementations of new functionalities.  Therefore, the developers can focus on the features to develop.

\subsection{Enabling and Disabling Modules}
Table~\ref{moudule_maintenance} shows the results when our proposed system is executed with the modulator compared to the existing system. As a test scenario, we activated the existing 12 modules at first. Then, we executed an addition (+number) and removal (-number) of modules in consecutive order in the existing system. Also, we have tested plug-in and plug-out facilities by setting configuration file to add and remove the modules  in due order in our state-of-the-art system.

From our experiment, the existing system does not provide any safety equipment \cite{7965323} in case that developers have to disable or enable their inspection modules. As result of that, it frequently generates code errors, duplicated codes, unnecessary maintenance cost of source code, and difficult handling of the modules due to no configuration  setting. However, our system supports a configuration file to avoid the situation that administrator has to know internal operations of the source code in.

\begin{table}[htbp]
\caption{Module maintenance comparison}
\begin{center}
\begin{tabular}{llll}
\hline
Step & Activity & \begin{tabular}[c]{@{}l@{}}Without modulator\\ (Jenkins)\end{tabular}     & 
\begin{tabular}[c]{@{}l@{}}With modulator\\ (TAOS-CI)\end{tabular} \\
\hline
1    & +12      & add   & plug-in                          \\
2    & -2       & remove& plug-out                         \\
3    & +1       & add   & plug-in                          \\
4    & -1       & remove& plug-out                         \\
5    & -1       & remove& plug-out                         \\
6    & -1       & remove& plug-out                         \\
7    & +2       & add   & plug-in                          \\
8    & -1       & remove& plug-out                         \\
9    & -1       & remove& plug-out                         \\
10   & -1       & remove& plug-out                         \\
11   & +1       & add   & plug-in                          \\
12   & -1       & remove& plug-out                         \\
\hline
     & Result   & \begin{tabular}[c]{@{}l@{}}code errors\\ code duplication\\ management cost\\ no configuration\end{tabular} & -  \\
\hline
\end{tabular}

\label{moudule_maintenance}
\end{center}
\end{table}

%% file: subsection/070_related_work.tex
  
\textit{Zhao et al.} \cite{8115619} suggest a more nuanced picture of how GitHub teams are adapting to, and benefiting from, continuous integration technology than suggested by prior work. Also, he empirically studied the impact of adopting TRAVIS CI \cite{6802994} on development practices in a collection of GitHub \cite{Vasilescu:2015:QPO:2786805.2786850} projects, and surveyed the developers responsible for introducing TRAVIS in those projects. However, this paper does not handle how to accelerate an integration speed of increased Pull Requests.

\textit{Wait For It} \cite{7180096}  describes on a quantitative study that tries to resolve which factors affect pull request \cite{7194588} evaluation latency in GitHub. Their preliminary models show that pull request review latency \cite{6976151} is complex, and depends on many predictors. They only concentrate on the regression modeling to fine reasons of the pull request evaluation latency in GitHub.

\textit{Wang et al.} \cite{4814157} depict a novel approach to assist \texttt{triagers} in detecting duplicate bug reports. In his approach, when a new bug report \cite{5487527} arrives, its natural language information and execution information are compared with those of the existing bug reports. The \texttt{triagers} mean people with the “Developer” role on the issue tracker. The \texttt{triagers} examine the suggested bug reports \cite{Dabbish:2012:SCG:2145204.2145396} to determine whether the new bug report duplicates an existing bug report. However, this paper needs to develop more-suitable evaluation techniques with practical situations.

\textit{Pham et al.} \cite{6606557} present insights about the contribution process on GitHub \cite{Nagappan:2008:IOS:1368088.1368160}, \cite{Tsay:2014:LTE:2635868.2635882} and shows how project owners assess pull requests with regard to testing. This paper is an exploratory first step to get an understanding of the testing norms, challenges, and strategies on social coding sites. Their research identified current challenges and solutions that are used in commercial and open source software development. However, they do not handle an advanced GitHub system architecture to improve the contribution process on GitHub.

%% file: subsection/080_conclusion.tex
As the complexity of the software gets worse, the software source code is repeatedly lagging and delayed. Especially, this problem causes a delay in firmware update of IoT devices in the edge computing environment. The existing systems are focused on the infrastructure for continuous integration. However, the proposed system focuses on the functionality of the source code management techniques rather than the scalability of the infrastructure for the fast continuous integration. The functionality research is very important to minimize problems that impede continuous integration of software platform.

We also propose a system mechanism that can reuse source code inspectors in most of the projects hosted on a source code management server with only the modification of the configuration setting. Also, if necessary, additional checking functions can be modularly developed. Moreover, we introduce the code coordination system that source code can be lightly and quickly re-assembled and managed.

%% file: subsection/090_acknowledgment.tex
We would like to thank the anonymous reviewers for their valuable feedback on earlier drafts of this paper. We gratefully acknowledge Hong Seok Kim and Gary Geunbae Lee for their feedback and comments, which helped to improve the content of this paper.

%% file: subsection/095_reference.tex
\bibliographystyle{IEEEtran} 


\bibliography{icce2019}